\documentclass{article}
\usepackage{spconf,amsmath,graphicx}

\usepackage{cite}
\usepackage{amsmath,amssymb,amsfonts}
\usepackage{algorithmic}
\usepackage{textcomp}
\usepackage{booktabs} 
\usepackage{lineno,hyperref}
\usepackage{caption}

\usepackage[dvipsnames]{xcolor}

\newcommand{\mbf}{\mathbf}

\usepackage{float}
\usepackage{gensymb}
\usepackage{etoolbox}
\usepackage{xspace}

\makeatletter
\DeclareRobustCommand\onedot{\futurelet\@let@token\@onedot}
\def\@onedot{\ifx\@let@token.\else.\null\fi\xspace}

\def\ie{\emph{i.e}\onedot}

\def\etal{\emph{et al}\onedot}
\makeatother

\title{AUGMENTING TRANSFERRED REPRESENTATIONS FOR STOCK CLASSIFICATION}

\name{Elizabeth Fons$^{1}$\thanks{This work was supported by the European Union's Horizon 2020 research and innovation programme under the Marie Sklodowska-Curie Grant Agreement no. 675044 (\url{http://bigdatafinance.eu/}), Training for Big Data in Financial Research and Risk Management. A. Iosifidis acknowledges funding from the Independent Research Fund Denmark project DISPA (Project Number: 9041-00004).} \qquad Paula Dawson$^{2}$ \qquad Xiao-jun Zeng$^{1}$ \qquad John Keane$^{1}$ \qquad Alexandros Iosifidis$^{3}$}

\address{$^{1}$ School of Computer Science, University of Manchester, UK.\\ 
      $^{2}$AllianceBernstein, London, UK.\\ 
      $^{3}$Department of Electrical and Computer Engineering, Aarhus University, Denmark.}
  
\begin{document}
\ninept
\maketitle
\begin{abstract}

Stock classification is a challenging task due to  high levels of noise and volatility of stocks returns. In this paper we show that using transfer learning can help with this task, by pre-training a model to extract universal features on the full universe of stocks of the S$\&$P500 index and then transferring it to another model to directly learn a trading rule. Transferred models present more than double the risk-adjusted returns than their counterparts trained from zero.
In addition, we propose the use of data augmentation on the feature space defined as the output of a pre-trained model (\ie augmenting the aggregated time-series representation). We compare this augmentation approach with the standard one, i.e. augmenting the time-series in the input space. We show that augmentation methods on the feature space leads to $20\%$ increase in risk-adjusted return compared to a model trained with transfer learning but without augmentation. 
\end{abstract}
\section{Introduction}

Stock market prediction is a challenging task primarily driven by a high degree of noise and volatility influenced by external factors such as extreme macroeconomic conditions, heightened correlations across multiple markets, and investor's behaviour. While much work has been done on stock movement prediction~\cite{Zhang2018, Tran2019, Passalis2018}, it remains an open research challenge. For the past decade, deep neural networks have exhibited very good performance in many different fields such as computer vision, natural language processing and robotics. 
In recent years, it has been shown that deep learning is capable of identifying nonlinearities in time series data leading to profitable strategies~\cite{Krauss2017, Fischer2018, Sethi2014}. 

A common approach when using deep learning for stock classification is to focus on developing a model that predicts the movement of an index or of stocks, and then build a simple trading rule in order to test the method's profitability (in some cases not taking into account transaction costs that could potentially erode some or all earnings)~\cite{Krauss2017, Sezer2018, Bao2017, Jiang2020}. Further, little work has been done using large-scale datasets such as all S$\&$P500 constituents in a survivorship bias-free way - survivorship bias is the tendency to view performance of existing stocks as a representative sample without regarding those that have gone bankrupt and leads to overestimation of historical performance~\cite{Garcia52}. In general, previous work has focused on  predicting either movement of an index or of a small number of stocks~\cite{Fischer2018}.

In this work, we present a model that can learn a trading rule directly from a large-scale stock dataset. For this, we propose using transfer learning, where we pre-train a model with past returns of all constituent stocks of the S$\&$P500 index, and then transfer it and fine-tune it on a dataset that has the trading rule included. Transfer learning is widely used in computer vision when the target dataset contains insufficient labeled data~\cite{Shaha2018, Yosinski2014}; to our knowledge, it has been rarely used in deep learning models for time series data~\cite{Fawaz2018}. Current approaches in trading strategies treat each market or asset in isolation, with few use cases of transfer learning in the financial literature~\cite{Koshiyama2020, Zhang2018}.

Motivated by increasing interest in data augmentation for time series, we propose using data augmentation to improve generalisation~\cite{Iwana2020, Wen2020}. Given that we are using transfer learning, we propose transforming the vector representation of data within the learned feature space (\ie augmenting the aggregated time series representation obtained from the output of the pre-trained model, following DeVries~\etal~\cite{devries2017}) and compare this with standard input space augmentation, done by applying transformations such as time warp, jittering, etc.~\cite{Iwana2020b, Um2017}.

The contributions of the paper are as follows:
\begin{itemize}
    \item We pre-train a model using all the constituents of the S$\&$P500 index to extract universal features and then we transfer this model to another to learn a trading rule. 
    \item We propose the use of data augmentation on the feature space defined as the output of the pre-trained model and we compare this approach with the standard augmentation in the input space. 
    \item We test our model by building the learned trading rule and calculate profitability taking into account transaction fees.
\end{itemize}

\section{Methodology}
\subsection{Dataset}
\begin{figure}
    \centering
    \includegraphics[width=\linewidth]{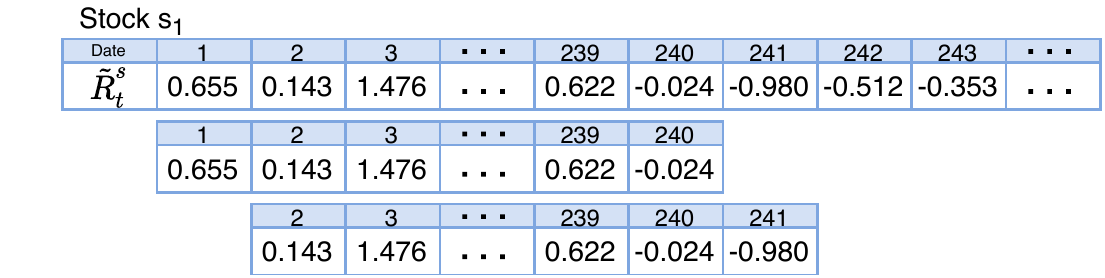}
    \caption{Construction of input sequences, segmented in $240$ time steps, with a moving window of one day.}
    \label{fig:data_diagram}
\end{figure}

The data used in this study consists of the daily returns of all constituent stocks of the S$\&$P500 index, from $1990$ to $2018$. It comprises $7000$ trading days, and approximately $500$ stocks per day. We use the data pre-processing scheme proposed by Krauss~\etal~\cite{Krauss2017}, where the data is divided into splits of $1000$ days, with a sliding window of $250$ days. Each split overlaps with the previous one by $750$ points, resulting in $25$ splits in total; a model is trained on each split. Inside each of the 25 splits, the data is segmented into sequences consisting on $240$ time steps $\{\tilde{R}^{s}_{t-239}, \ldots, \tilde{R}^{s}_{t}\}$ for each stock $s$, with a sliding window of one day, as shown in Figure~\ref{fig:data_diagram}. The first $750$ days consist of the training set, and the test set is formed by the last $250$ days. This leads to a training with approximately 255K samples ((750-240)*500) and the test set with approximately 125K samples.
The data is standardised by subtracting the mean of the training set ($\mu_{train}$) and dividing by the standard deviation ($\sigma_{train}$):
$\tilde{R}^{s}_t = \frac{R^{s}_t - \mu_{train}}{\sigma_{train}}$, with $R^s_t$ being the return of stock $s$ at time $t$. 
\subsection{Trading rule and training targets}
A goal of this work is to evaluate whether training a neural network by directly passing the trading rule as the classification target is better than training a binary classifier and then applying a trading rule. Therefore, to construct the target, taking the $\approx500$ daily stocks, we ranked them by their returns and the top $K$ are labeled as {\it buy}, the bottom $K$ are labeled as {\it sell}, and the rest as {\it do nothing}. Following the recommendation from Krauss~\etal~\cite{Krauss2017}, we use $K=10$. This leads to a highly imbalanced dataset, with a label proportion of $10:10:480$. 

The source network is trained as a binary classification task, where the target variable $Y^s_{t+1}$ for stock $s$ and date $t$ can take to values, 1 if the returns are above the daily median (trend up) and 0 if returns are below the daily median (trend down).

\subsection{Architecture and training}
The network architecture selected for transfer learning (source network) is a one layer LSTM proposed by Krauss~\etal~\cite{Krauss2017}. We chose this architecture because it achieved good performance on a large, liquid dataset - with similarities to the S$\&$P500. The network is a single layer LSTM with $25$ neurons, and a fully connected two-neuron output. We use a learning rate of $0.001$, batch size $128$ and early stopping with patience $10$ with RMSProp as optimizer. We implemented the neural network architectures using pytorch~\cite{pytorch} in Python.

After training the source network on the 25 splits of data, we have 25 neural networks. We then remove the output layer and replace it with a fully connected layer of $n$ neurons and an output layer of 3 neurons. The weights of the LSTM layer are fixed (there is no retraining), and the fully connected layer and output are trained with the new target data that incorporates the trading rule. Fawaz~\etal~\cite{Fawaz2018} proposes fine-tuning the transferred model parameters on the new dataset, but this leads to poor performance in our case, so the weights on the LSTM are left fixed, and just the new added layers are trained. 

The loss used for training in all cases is the cross-entropy loss. An alternative approach is to incorporate a loss term that takes into account the direct information on the positions from the network, and optimizes the average return, as follows:
\begin{equation}
    \mathcal{L}_{R+CE}(\mbf{\Theta)} = \mathcal{L}_{CE} + \alpha \mathcal{L}_{returns} = \mathcal{L}_{CE} + - \alpha \frac{1}{B} \sum R(i,t) 
    \label{eq:loss}
\end{equation}
where $B$ is the size of the batch, $R(i,t)$ is the return captured by the network prediction for asset $i$ at time $t$ and $\alpha$ is a scaling factor so both terms are equally represented. 

As the dataset is highly imbalanced, we sample the training data with higher probability in the minority classes, obtaining a balanced representation of the classes in each batch.
%
\subsection{Augmentation}
\begin{figure*}
    \centering
    \includegraphics[width=0.95\linewidth]{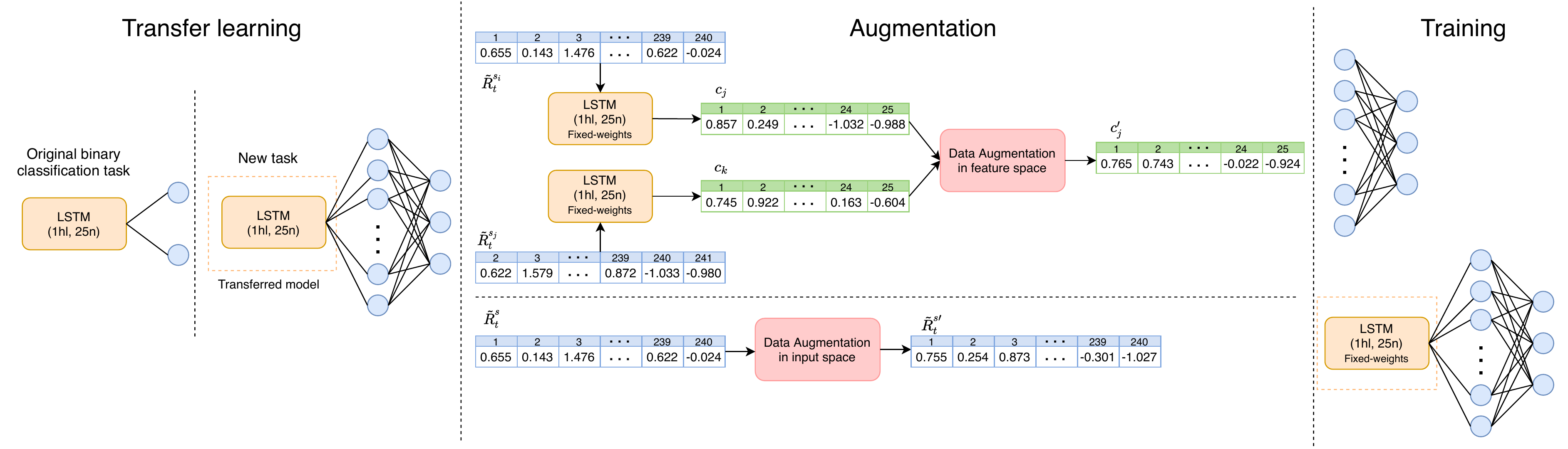}
    \caption{General training process with transfer learning and data augmentation. The panel on the left shows the pre-trained model and the new transferred one. On the centre, workflow of the two augmentation approaches, data augmentation on the feature space (top) and data augmentation on the input space (bottom; on the right, the resulting networks to be trained.}
    \label{fig:da_scheme}
\end{figure*}

The approximately 255K samples of the training set are divided into training and validation with a proportion $80/20$. The validation set is used for early stopping the training. Each train set is augmented one time, \ie we apply an augmentation method to the training data and the final set corresponds to the original data plus the augmented one, doubling the amount of samples. We study two forms of data augmentation: applying random transformations on the input data and doing data augmentation on the feature vector obtained by evaluating the input on the fixed LSTM layer. Figure \ref{fig:da_scheme} shows the training process with transfer learning and both forms of data augmentation.

\subsubsection{Data augmentation in feature space}
\begin{figure}
    \centering
    \raisebox{2em}{\tiny a)}\includegraphics[width=0.98\linewidth]{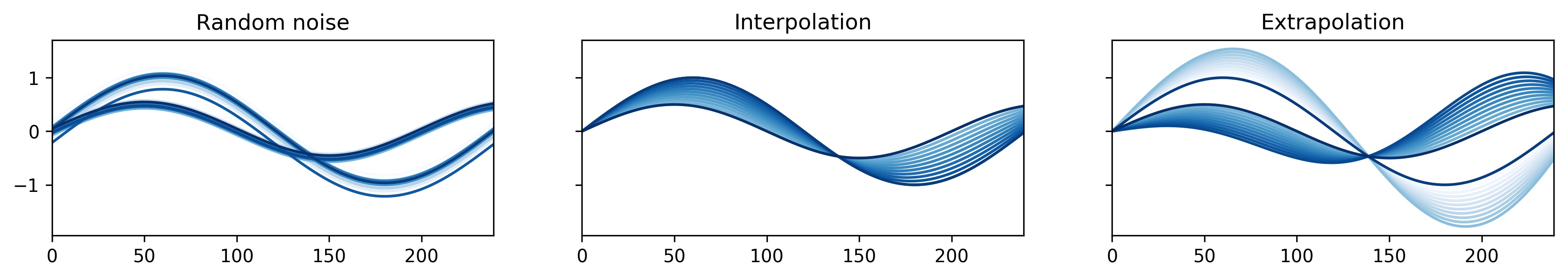}\\
    \raisebox{4em}{\tiny b)}\includegraphics[width=0.99\linewidth]{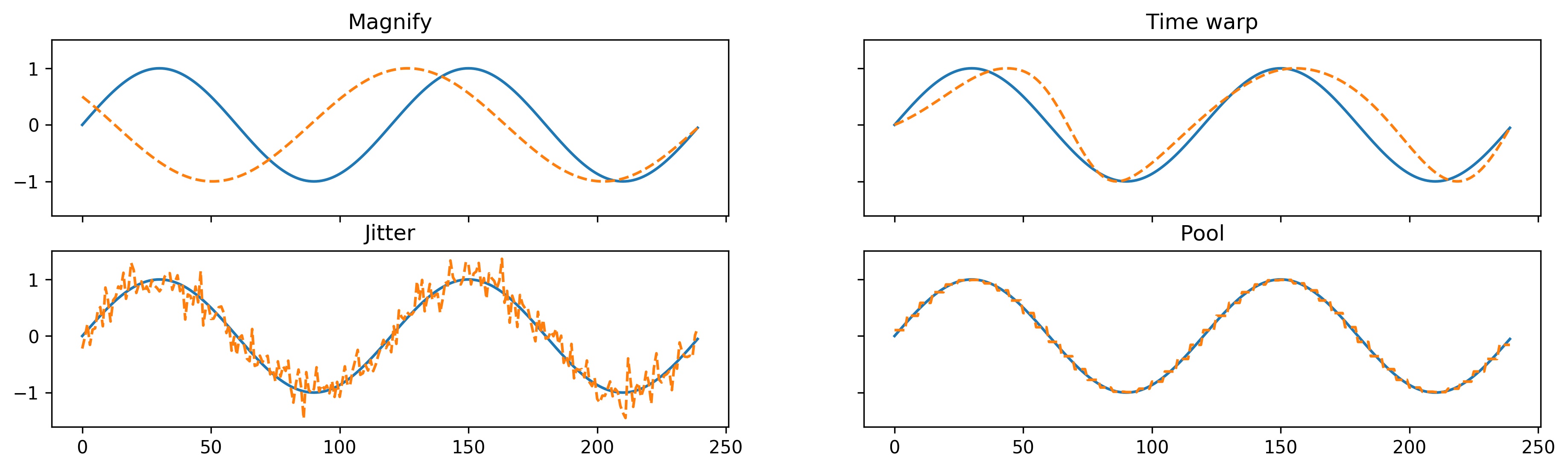}
    \caption{a) Examples of feature space augmentation methods on a sine wave. Random noise added with $\gamma=0.3$, interpolation and extrapolation  between two sinusoids for values of $\lambda$ between 0 and 1. b) Examples of input augmentation methods on a sine wave. Blue line corresponds to the original series and the dotted orange lines correspond to the augmented pattern.}
    \label{fig:da_latent}
\end{figure}

Data augmentation in the feature space as proposed by DeVries~\etal~\cite{devries2017} allows new samples to be generated in a domain-agnostic approach. It is an intuitive and direct approach where each sample is projected into feature space by feeding it through the fixed LSTM layer and then applying a transformation to the transformed vector (sometimes called context vector). The methods used to augment data in the feature space are described as follows and an example is shown in Figure \ref{fig:da_latent}a:\\
{\it Interpolation:} for each sample in the dataset, we find its $K$ intra-class nearest neighbours in feature space. For each pair of neighbouring vectors, a new vector is generated using interpolation:
\begin{equation*}
    \mathbf{c}_j^\prime = (\mathbf{c}_k - \mathbf{c}_j) \lambda + \mathbf{c}_j
\end{equation*}
where $\mathbf{c}_j^\prime$ is the new vector, $\mathbf{c}_k$ and  $\mathbf{c}_j$ are neighboring vectors and $\lambda$ is a variable in the range $\{0,1\}$. We used $\lambda = 0.2$.\\
{\it Extrapolation:} similarly, we apply extrapolation to the feature space vectors in the following way:\\
\begin{equation*}
    \mathbf{c}_j^\prime = (\mathbf{c}_j - \mathbf{c}_k) \lambda + \mathbf{c}_j
\end{equation*}
In this case, $\lambda$ is in the range $\{0,\infty\}$. We used $\lambda = 0.2$.\\
{\it Random noise (noise):} Gaussian noise is generated with zero mean and per-element standard deviation calculated across all transformed vectors in the dataset; the noise is scaled by a global parameter $\gamma$:
\begin{equation*}
        \mathbf{c}_i^\prime = \mathbf{c}_i + \gamma X, X \sim \mathcal{N}\{0, \sigma_i^2\}
\end{equation*}
{\it Jittering (Jit-feat):} Random noise with mean $\mu = 0$ and standard deviation $\sigma = 0.05$ is added to the context vector.

\subsubsection{Data augmentation in input space}
Most cases of time series data augmentation correspond to random transformations in the magnitude and time domain, such as jittering (adding noise), slicing, permutation (rearranging slices) and magnitude warping (smooth element-wise magnitude change). The following methods were used for evaluation and are shown in Figure~\ref{fig:da_latent}b.

{\it Magnify:} a variation of window slicing proposed by Le Guennec et al~\cite{guennec2016}. In window slicing, a window of $90\%$ of the original time series is selected at random. Instead, here we randomly slice windows between $40\%$ and $80\%$ of the original time series, but always from the fixed end of the time series (\ie we slice the beginning of the time series by a random factor). Randomly selecting the starting point of the slicing would make sense in an anomaly detection framework, but not on a trend prediction problem as is our case. The resulting time series is interpolated to the original size.

{\it Jittering (Jit-inp):} Gaussian noise with a mean $\mu = 0$ and standard deviation $\sigma = 0.05$ is added to the time series \cite{Um2017}. This is analogous to the random noise method applied to the context vectors.

{\it Pool:} Reduces the temporal resolution without changing the length of the time series by averaging a pooling window. We use a window of size $3$. This method is inspired by the resizing data augmentation process followed in computer vision.

\section{Evaluation}
\subsection{Transfer learning}
We test transfer learning on two networks, one with a fully connected layer of $25$ neurons and one with $100$ neurons, and compare the proposed networks trained using transfer learning with the same topology trained from scratch. 
In order to avoid random initialization conflicts on the non-transferred networks, we train three separate instances with different initial weight values (\ie we use three different seeds on the random initialisation) and average their performance. 

Given that we want to evaluate financial performance, we build a portfolio by ranking the output of the networks labeled class 1 ({\it buy}) and 2 ({\it sell}); we then take the 10 with highest probability for each class and build a long-short portfolio. 
Portfolios are analysed after transaction costs of 5bps per trade. 
The portfolio performance metric we use is Information ratio (IR) - the ratio between excess return (portfolio returns minus benchmark returns) and tracking error (standard deviation of excess returns) \cite{IR}. As the portfolios are long-short, they are market-neutral, therefore, performance of the portfolio in independent of performance of the market and no benchmark has to be subtracted. 
We also calculate the downside information ratio - the ratio between excess return and the downside risk (variability of under-performance below the benchmark), that differentiates harmful volatility from total overall volatility. We compare our results with Krauss~\etal~\cite{Krauss2017}, in which a binary classifier is trained and then the trading rule is applied. We also calculate two classification metrics, accuracy and macro-F1 expressed by the mean and standard deviation over the 25 data splits. 
\begin{table*}[]
    \caption{Performance of the $k=10$ long-short portfolios after transaction costs, built from models trained from zero and models whose weights where transferred from a pre-trained model.}
    \centering
    \resizebox{0.65\linewidth}{!}{
\begin{tabular}{lccccccc}
\toprule
Method                              & Ann ret       & Ann vol & IR &  D. Risk & DIR & Acc   & Macro-F1  \\
\midrule
LSTM~\cite{Krauss2017}               &  $29.2$       &   $28.66$     &  $1.02$  &    $19.08$ &      $1.53$ & --- & --- \\
No TL (25)+$\mathcal{L}_{CE}$       &  12.99        &   38.15       &  0.34  &  25.48 &  0.52 &   73.13$\pm$18.94 &   $\mbf{33.57\pm6.5}$  \\
No TL (25)+$\mathcal{L}_{R+CE}$     &  19.58        &    38.9       &  0.51  &   25.9 &   0.77  &  59.64$\pm$20.48 &  29.21$\pm$6.98   \\
TL+FC(25)+$\mathcal{L}_{CE}$        &  $32.25$      &   30.29       &  1.06  &  19.6  &  1.65 &    68.34$\pm$16.5 & 31.79$\pm$5.12 \\
TL+FC(25)+$\mathcal{L}_{R+CE}$      & $\mbf{34.62}$ &    30.2       &  {\bf 1.15}  &  19.59 &   {\bf 1.77}  &  64.79$\pm$16.86 &  30.72$\pm$5.28   \\
No TL (100)+$\mathcal{L}_{CE}$      &  5.05         &   42.60       &  0.12  &  29.28 &   0.19 &  $\mbf{74.69\pm21.14}$ &   33.37$\pm$6.94 \\
No TL (100)+$\mathcal{L}_{R+CE}$    &  21.05        &   39.95       &  0.55  &    25.9 &   0.84  &   $57.02 \pm 21.95$ & $28.24 \pm 8.12$ \\
TL+FC(100)+$\mathcal{L}_{CE}$       &  30.83        &   30.31       &  1.02  &    19.79 &   1.56 &  68.88$\pm$15.93 &   31.95$\pm$4.76  \\
TL+FC(100)+$\mathcal{L}_{R+CE}$     &  32.14        &   29.97       &  1.07  &  19.87 &   1.62  &  64.72$\pm$17.25 &   30.7$\pm$5.41  \\
\bottomrule
\end{tabular}}
    \label{tab:tfl_resample}
\end{table*}

Table~\ref{tab:tfl_resample} shows the performance of the models trained using only the cross-entropy loss with and without transfer learning, as well as training with the combined loss of cross-entropy and return maximization from equation \ref{eq:loss}. We  see that, in all cases, the models trained from scratch show a poor financial performance as shown by the information ratio. All transferred learned models show an equal or higher performance than the baseline LSTM method, with the models trained with the combined loss having a slightly higher performance. The classification metrics are higher for models without transfer learning and trained only on the cross-entropy loss. Using the combined loss hurts classification metrics in all cases, but as expected, it improves portfolio performance.

\subsection{Data augmentation}
Table~\ref{tab:aug_25_resample} shows the performance of augmentation methods for the architectures with a fully connected layer of 25 neurons and table~\ref{tab:aug_100_resample} models with a fully connected layer of 100 neurons. All models were trained using the combined loss of equation \ref{eq:loss}, unless stated otherwise.
In both topologies, augmentation on the input space is ineffective, and in most cases it decreases performance with respect to the model trained with transfer learning but without augmentation. In contrast, models trained with augmentation on the feature space tend to improve on IR, in the case of jittering.
\begin{table*}
\begin{minipage}[t]{0.5\textwidth}
\centering
\captionof{table}{Performance of the $k=10$ long-short portfolios after transaction costs, for the TL+FC(25) model trained with different augmentation methods and the combined loss $\mathcal{L}_{R+CE}$.}
\resizebox{\linewidth}{!}{
\begin{tabular}{lccccccc}
\toprule
Method                          & Ann ret & Ann vol & IR     &D. Risk &    DIR  & Acc   & Macro-F1  \\
\midrule
LSTM\cite{Krauss2017}           &    29.2 &   {\bf 28.66} &  1.02  &  19.08 &   1.53  & ---               & ---       \\
No TL (25)+$\mathcal{L}_{CE}$       &  12.99        &   38.15       &  0.34  &  25.48 &  0.52 &   73.13$\pm$18.94 &   $\mbf{33.57\pm6.5}$  \\
TL+FC(25)+$\mathcal{L}_{CE}$        &  $32.25$      &   30.29       &  1.06  &  19.6  &  1.65 &    68.34$\pm$16.5 & 31.79$\pm$5.12 \\
TL+FC(25)+$\mathcal{L}_{R+CE}$  &  34.62  &   30.20 &  1.15  &  19.59 &   1.77  &  64.79$\pm$16.86      &   30.72$\pm$5.28          \\
\midrule
TL+FC(25) Extrapolation          &   {\bf 39.70} &   29.43 &   {\bf 1.35}  &  18.96 &   {\bf 2.09}  &  62.90$\pm$17.87 &    30.10$\pm$5.81  \\
TL+FC(25) Interpolation         &   36.87 &   29.69 &   1.24  &  {\bf 18.93} &   1.95  &  62.46$\pm$17.80 &    29.95$\pm$5.74  \\
TL+FC(25) Noise            &   30.97 &   29.15 &   1.06  &  19.14 &   1.62  &  62.43$\pm$18.12 &   29.95$\pm$5.81   \\
TL+FC(25) Jitter-feat            &   39.11 &   29.93 &   1.31  &  19.22 &   2.03  &  62.71$\pm$17.84 &   30.04$\pm$5.71   \\
\midrule
TL+FC(25) Jitter-input                &   29.74 &   39.94 &  0.96  &  20.12 &   1.48  &  68.23$\pm$16.62  &   31.75$\pm$5.06   \\
TL+FC(25) Magnify               &   20.39 &   29.41 &  0.69  &  19.86 &   1.03  &  63.78$\pm$16.78  &   30.42$\pm$5.47   \\
TL+FC(25) Pool                  &   27.18 &   29.96 &  0.91  &  19.64 &   1.38  &  57.71$\pm$17.43  &   28.38$\pm$5.72    \\
TL+FC(25) Time Warp                    &   32.76 &   29.46 &  1.11  &  19.21 &   1.71  &  61.81$\pm$19.96  &   29.80$\pm$5.48   \\
\bottomrule
\end{tabular}
}
\label{tab:aug_25_resample}
\end{minipage}~
\begin{minipage}[t]{0.5\textwidth}
\centering
\captionof{table}{Performance of the $k=10$ long-short portfolios after transaction costs, for the TL+FC(100) model trained with different augmentation methods and the combined loss $\mathcal{L}_{R+CE}$.}
\resizebox{\linewidth}{!}{
\begin{tabular}{lccccccc}
\toprule
Method                      & Ann ret & Ann vol & IR     &D. Risk &    DIR  & Acc   & Macro-F1  \\
\midrule
LSTM\cite{Krauss2017}       &    29.2 &   28.66 &  1.02  &  19.08 &   1.53  & ---               &  ---               \\
No TL (100)+$\mathcal{L}_{R+CE}$    &  21.05        &   39.95       &  0.55  &    25.9 &   0.84  &   57.02$\pm$21.95 & 28.24$\pm$8.12 \\
TL+FC(100)+$\mathcal{L}_{CE}$       &  30.83        &   30.31       &  1.02  &    19.79 &   1.56 &  68.88$\pm$15.93 &   31.95$\pm$4.76  \\
TL+FC(100)+$\mathcal{L}_{R+CE}$     &  32.14        &   29.97       &  1.07  &  19.87 &   1.62  &  64.72$\pm$17.25 &   30.7$\pm$5.41  \\
\midrule
TL+FC(100) Extrapolation   &   27.38 &   29.33 &  0.93  &  19.42 &   1.41   &  62.74$\pm$17.73 &  30.05$\pm$5.81 \\
TL+FC(100) Interpolation    &   30.84 &   29.72 &  1.04  &  19.38 &   1.59   &  62.49$\pm$17.35 &  30.01$\pm$5.63 \\
TL+FC(100) Noise       &   29.02 &   29.44 &  0.99  &   19.2 &   1.51   &  62.2$\pm$17.82  &  29.87$\pm$5.73 \\
TL+FC(100) Jitter-feat      &   {\bf 37.14} &   29.31 &  {\bf 1.27}  &  {\bf 18.92} &   {\bf 1.96}   &  61.84$\pm$17.89 &  29.75$\pm$5.75 \\
\midrule
TL+FC(100) Jitter-input           &   29.49 &   30.32 &  0.97  &  19.73 &   1.49   &  67.79$\pm$17.18 &  31.64$\pm$5.46  \\
TL+FC(100) Magnify          &   22.11 &   30.36 &  0.73  &  20.21 &   1.09   &  67.12$\pm$16.68 &  30.34$\pm$5.27 \\
TL+FC(100) Pool             &  27.64  & 29.50 &  0.94  & 18.98  &   1.46   &  57.64$\pm$17.89 &  28.37$\pm$5.89 \\
TL+FC(100) Time Warp         &   26.64 &   29.65 &  0.90  &  19.49 &   1.37   &  65.55$\pm$18.03 &  29.69$\pm$5.89 \\
\bottomrule
\end{tabular} 
}
\label{tab:aug_100_resample}
\end{minipage}
\end{table*}

\section{Conclusions}
In this work, we have shown that using transfer learning on a stock classification task where a trading rule is included in the training dataset improves financial performance when compared to training a neural network from scratch. To evidence this we built long-short portfolios following the proposed trading rule. Models trained with transfer learning improve information ratio by more than double with respect to models trained without a source model. We also showed that using a training loss that combines a classification objective with maximization of returns improves risk adjusted returns when compared with the single cross-entropy loss. 

Finally, we investigated the use of data augmentation on the feature space (defined as the output of the pre-trained model) and compared it with traditional data augmentation methods on the input space. Augmentation on the input space improves up to $\%20$ risk adjusted returns when compared to a transferred model without augmentation.  

%
\bibliographystyle{IEEEbib}
\bibliography{biblio}

\end{document}